\documentclass[12pt]{article}
\usepackage{amstex,epsfig}
\def\g5{\gamma^5}

\def\d4k{{d^4k\over (2\pi)^4}}

\newcommand{\beq}{\begin{eqnarray}}
\newcommand{\eeq}{\end{eqnarray}}
\newcommand{\beqno}{\begin{eqnarray*}}
\newcommand{\eeqno}{\end{eqnarray*}}
\def\lsim{\mathrel{\rlap{\lower4pt\hbox{\hskip1pt$\sim$}}
    \raise1pt\hbox{$<$}}}         
\def\gsim{\mathrel{\rlap{\lower4pt\hbox{\hskip1pt$\sim$}}
    \raise1pt\hbox{$>$}}}         

\begin{document}
%
\title{Peripheral Production of Sigmas in Proton Proton Collisions}

\author{Leonard S. Kisslinger\\
        Department of Physics,\\ \vspace{3mm}
       Carnegie Mellon University, Pittsburgh, PA 15213\\
       Wei-hsing Ma and Pengnian Shen\\
       Institute of High Energy Physics, Academia Sinica\\
       Beijing 100039, P. R. China}
\maketitle
\indent
\begin{abstract}
The Pomeron, which dominates high energy elastic and diffractive  
hadronic processes, must be largely gluonic in nature. We use a recent
picture of a scalar glueball/sigma system with coupling of the sigma
to glue determined from experiment to predict strong peripheral sigma
production seen in the p p $\pi^o\pi^o$ final state.
\end{abstract}

\vspace{0.5 in}

\noindent
PACS Indices: 12.40.Gg, 12.38.Lg, 13.85.Dz, 13.60.Le
\newpage
\section{Introduction}
\hspace{.5cm}

  Although all known hadrons seem to lie on Regge trajectories,
it has long been known that high energy elastic and diffractive processes
are not consistent with the meson trajectories \cite{pcol}.
The Pomeron, with the property that $\alpha_P$(0) $\simeq$ 1.0, dominates 
these high energy processes. Also peripheral processes, which correspond to
production of low momentum particles from the Regge trajectory, at high 
energies
are given by the emission of the peripheral particles from the Pomeron.
The most important feature of the Pomeron is that it must
be gluonic in nature. Thus peripheral production at high energy is given by
the coupling  of the peripheral particles to the gluonic field.
 
   In the present work we study sigma peripheral production in high emergy
proton-proton collisions.  It is based on the model\cite{lk1,lk2} that 
there exists a light scalar glueball strongly coupled to 
the I=0 two-pion system, which we call the glueball/sigma model,
and our recent work\cite{km} that this system might lie on the 
daughter trajectory of the pomeron.  The sigma/glueball model was recently 
proposed\cite{lk1,lk2} based on three observations: 1) at low energies the
the scalar-isoscalar $\pi-\pi$ sysem is observed 
in $\pi-\pi$ scattering\cite{zb} to be a Breit-Wigner 
resonance, which we call the sigma; 2) the sigma seems to dominate
scalar glueball decay\cite{bes}; and 3) in QCD sum rule calculations we
find\cite{lk1} a light scalar glueball far below the coupled scalar
glueball-meson systems which we find correspond to the f$_o$(1370) and
f$_o$(1500). Our proposed glueball/sigma resonance is a coupled-channel
glueball-2$\pi$ system with a mass and
width both about 400 MeV. With this picture it was predicted\cite{lk2}
that there will be found a large branching ratio for the decay of the
P$_{11}$(1440) baryon resonance to a sigma and a nucleon.

  In our earlier work\cite{km} we showed that the coupling of the 
pomeron to the nucleon can be predicted using the glueball sigma model
with no free parameters, and that this coupling agrees within expected
errors with a phenomenological 
Pomeron exchange model\cite{dl} that is consistent with many high energy
experiments. Also, recent work suggests\cite{lm} that the $\xi(2230)$, if it 
turns out to be a tensor glueball, might lie on the Pomeron itself.
The present work, however does not make use of any model of the Pomeron. It 
only uses the fact that the pomeron is gluonic in nature and the 
glueball/sigma picture of Ref \cite{lk2}. From this we derive the cross 
section for peripheral production of $\pi^o\pi^o$ through the 
glueball/sigma resonance in high energy pp scattering.  We find a large 
branching ratio that can be tested in experiment.

\section{Sigma Peripheral Production}
\hspace{.5cm}

  As depicted in Fig. 1a, the form of the elastic proton-proton scattering 
amplitiude with Pomeron exchange is given by
\beq
     A^{pp} & = & V(t) D^P(t,s) V(t),
\label{1}
\eeq
where V is the vertex function given by the Pomeron residue and D$^P$ is
the Pomeron propagator.  We write this propagator as
\beq
        D^P(q) & = & \int d^4x e^{iq\cdot x}<0|T[[G(x)G(x)][G(0)G(0)]]|0>,
\label{2}
\eeq
where [G(x)G(x)] is a symbolic form for the current of the Pomeron.
With a similar notation the amplitude for the peripheral
production of $\sigma \rightarrow \pi\pi$ as shown in Fig. 1b is given by
\begin{figure}
\begin{center}
\epsfig{file=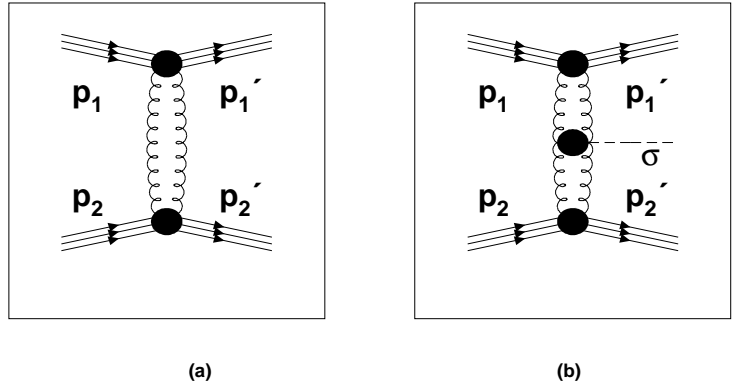,width=12cm}
\caption{a)Elastic p-p scattering, b) peripheral production with Pomeron 
exchange.}
{\label{Fig.1}}
\end{center}
\end{figure}
\beq
     A^{pp\sigma} & = & V(t) D^P_\sigma(t,s) V(t),
\label{3}
\eeq
where D$^P_\sigma$ is the propagator of the exchanged Pomeron coupled to a
$\sigma$, which decays to the I=0 2$\pi$ resonance. 
Of course, the cross section of the diffractive producton process depends 
on the momentum transfers to the two interacting nucleons, $t_{1}$ and 
$t_{2}$, which are different from the momentum transfer, t, for elastic  
scattering. However, since we are discussing sigma peripheral production, 
the sigma meson carries a very small momentm momentum $p_{\sigma}$. Therefore 
to a very good approximation it follows that $t_{1} \simeq t_{2} \simeq t$.  
Note that our calculations are for 50 GeV protons and the sigma momentum is 
of the order of 0.3 GeV.

Using the external field
method with the sigma treated as an external field we write this propagator as
\beq
D^P_\sigma(q) & = & \int d^4x 
e^{iq\cdot x}<0|T[[G(x)G(x)][G(0)G(0)]]_\sigma|0>. 
\label{4}
\eeq
Assuming factorization, we use $ [[G(x)G(x)][G(0)G(0)]]_\sigma \simeq
[G(x)G(x)][G(0)G(0)]_\sigma$. For low-momentum sigmas we neglect
the additional form factor for the sigma-gluonic coupling, since it is
approximately unity.  This gives 
\beq
           D^P_\sigma(t,p_\sigma) & = & g_\sigma G_\sigma(p_\sigma) D^P(t).
\label{5}
\eeq
In Eq.(\ref{5}) g$_\sigma$ is the $\sigma$-gluon coupling constant derived
in Ref \cite{lk2} and G$_\sigma$ is the Breit-Wigner resonance propagator
of the sigma.  Introducing the appropriate phase space factors we find 
from Eqs.(\ref{1}, \ref{3}, and \ref{5}) the relationship
\beq
      \frac{d^2 \sigma^{pp\sigma}/dt dE_\sigma}{d\sigma^{pp}/dt}
 & = &
\frac{g_\sigma^2 p_\sigma}{2 \pi^2(p_\sigma^4 + M_\sigma^2 \Gamma_\sigma^2)},
\label{6}
\eeq
where we have included the fact that both gluons can radiate sigmas.
The results for the ratio of cross sections in the center of mass frame are 
shown in Fig. 2.
\begin{figure}
\begin{center}
\epsfig{file=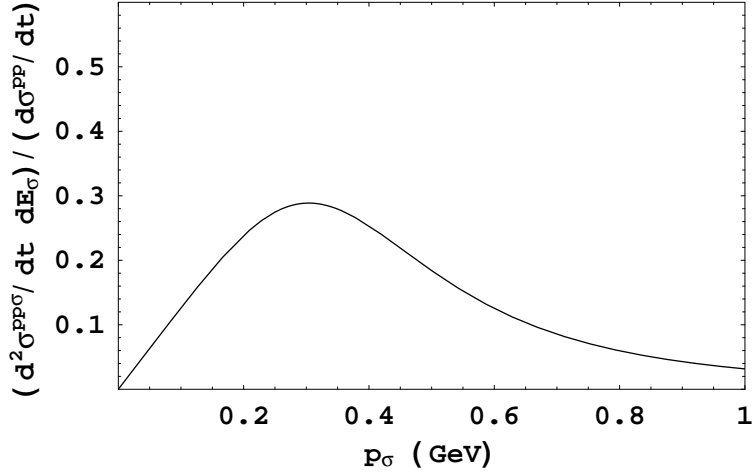,width=10cm}
\caption{Ratio of differential cross sections of $pp \rightarrow pp\sigma$
process to elastic $pp \rightarrow pp$ process.}
{\label{Fig.2}}
\end{center}
\end{figure}

For comparison with experiment we rewrite our results in terms of
the $\sigma$ rapidity distribution which is a function
of the transverse momentum of the $\sigma$. With the standard definition
the rapidity of the $\sigma$ and the $\sigma$ energy are given by 
\beq
\label{7}
           y & = & tanh^{-1}(\frac{p_{\sigma z}}{E_\sigma}), \\
          E_\sigma & = & \sqrt{m_\sigma^2 + p_{\sigma\perp}^2}cosh(y),
\eeq
where $p_{\sigma\perp}^2  = \sqrt{p_x^2 + p_y^2}$ is the transverse 
momentum of the $\sigma$.
Using the approximation that the t variable is the same as for elastic
p-p scattering for peripheral production, as explained above, we integrate 
over the t variable to obtain from Eq.(\ref{6})
\beq
\label{8}
        \frac{d \sigma^{pp\sigma}}{dy} & = & \sigma^{pp}_{tot}
 \frac{g_\sigma^2}{2\pi^2}\frac{ \sqrt{(m_\sigma^2 + p_{\sigma\perp}^2)
 ((m_\sigma^2 + p_{\sigma\perp}^2)cosh^2y-m_\sigma^2)}}
 {((m_\sigma^2 + p_{\sigma\perp}^2)cosh^2y-m_\sigma^2)^2 +m_\sigma^2
 \Gamma_\sigma^2}sinh(y).
\eeq
Using the experimental fit to the total elastic p-p cross section at high
energy\cite{tot}, $\sigma^{pp}_{tot} = 21.70 s^{0.0808} + 56.08 s^{-0.4525}$
mb, which is similar to the fit obtained with the Pomeron model of 
Ref. \cite{dl}, 
we obtain the results shown in Fig. 3.        
\begin{figure}
\begin{center}
\epsfig{file=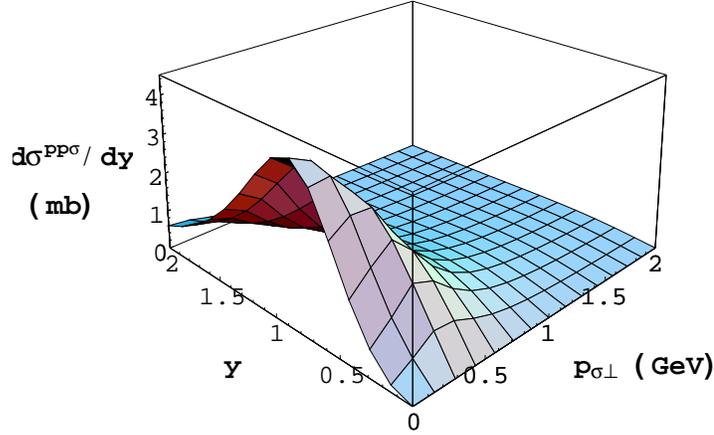,width=10cm}
\caption{Rapidity and transverse momentum distributiion for sigma production.}
{\label{Fig.3}}
\end{center}
\end{figure}

As seen in Fig. 3,
in our model the cross section for peripheral production near y = 1.0 and
low momentum transfer is quite large, about 2 mb.
Note that the charged $\pi^+\pi^-$ channel might not be as 
satisfactory because of the $\rho$-meson background from processes at the
nucleon vertices, which makes the interpretation of experiment more difficult.

\section{Conclusions}
\hspace{.5cm}

We have derived the cross section for diffractive sigma production in
high energy proton-proton collisions. By mapping out the low-energy
spectrum of the two $\pi_o$s one will find the sigma resonance if this
glueball/sigma model is correct. If so, the production of sigmas can be
used as a signal for glueballs and hybrids as well as the Pomeron.

The work was supported in part by the U.S. National Science Foundation
grants PHY-00070888, INT-9514190 and the National Science Foundation of
China grant 1977551B.

\end{document}